\documentclass{Interspeech}
\usepackage{booktabs}  
\usepackage{tabularx}  
\usepackage{array}  
\usepackage{makecell} % 让多行文本居中
\usepackage{multirow} % 确保支持 multirow
\interspeechcameraready

\title{Speaker Diarization with Overlapping Community Detection Using Graph Attention Networks and Label Propagation Algorithm}
\author[affiliation={1}]{Zhaoyang}{Li}
\author[affiliation={1}]{Jie}{Wang}
\author[affiliation={2}]{XiaoXiao}{Li}
\author[affiliation={1}]{Wangjie}{Li}
\author[affiliation={1}]{Longjie}{Luo}

\author[affiliation={1}]{Lin}{Li*}
\author[affiliation={3}]{Qingyang}{Hong*}

\affiliation{School of Electronic Science and Engineering}{Xiamen University}{China}
\affiliation{School of Electronic Information}{Beijing Jiaotong University}{China}
\affiliation{School of Informatics}{Xiamen University}{China}
\email{23120231150269@stu.xmu.edu.cn, \{lilin,qyhong\}@xmu.edu.cn}
\keywords{speaker diarization, graph attention network, label propagation, overlapping community detection}

\usepackage{comment}
\begin{document}

\maketitle

\renewcommand{\thefootnote}{\fnsymbol{footnote}}
\footnotetext[1]{Corresponding author.}
\renewcommand{\thefootnote}{\arabic{footnote}}

\begin{abstract}
In speaker diarization, traditional clustering-based methods remain widely used in real-world applications. However, these methods struggle with the complex distribution of speaker embeddings and overlapping speech segments. To address these limitations, we propose an Overlapping Community Detection method based on Graph Attention networks and the Label Propagation Algorithm (OCDGALP). The proposed framework comprises two key components: (1) a graph attention network that refines speaker embeddings and node connections by aggregating information from neighboring nodes, and (2) a label propagation algorithm that assigns multiple community labels to each node, enabling simultaneous clustering and overlapping community detection. Experimental results show that the proposed method significantly reduces the Diarization Error Rate (DER), achieving a state-of-the-art 15.94\% DER on the DIHARD-III dataset without oracle Voice Activity Detection (VAD), and an impressive 11.07\% with oracle VAD.

\end{abstract}

\section{Introduction}

Speaker diarization is designed to segment the conversational audio signals into distinct segments with labeled speakers' identities, effectively addressing the problem of ``who spoke when" \cite{park2022review}. In recent years, speaker diarization technology has gained widespread application across diverse fields, such as meeting transcription, conversational analysis \cite{yella2014artificial}. However, real-world scenarios still face issues like speech overlap and unknown speaker counts.

Current representative speaker diarization systems can be broadly categorized into two types: end-to-end systems and modular systems. End-to-end systems can directly process raw audio and produce diarization results \cite{fujita2019end,fujita2020neural}. These systems demonstrate notable strengths in handling overlapping speech, but still face challenges such as ambiguous speaker label assignments. Additionally, methods like Target Speaker Voice Activity Detection (TS-VAD) \cite{he2023ansd,cheng2023target} are highly reliant on initial clustering results.
In contrast, the modular framework, also known as Clustering-based Speaker Diarization (CSD), typically comprises four components: Voice Activity Detection (VAD), a speaker representation extractor, a clustering module, and a post-processing module. The clustering module often employs traditional unsupervised clustering methods such as Agglomerative Hierarchical Clustering (AHC) \cite{sell2016priors}, or Spectral Clustering (SC) \cite{wang2022similarity}. However, the complex distribution of speaker embeddings and the sensitivity of various methods to hyperparameters can impact the classification performance of these systems.

For high-dimensional speaker embeddings such as x-vector \cite{wang2018speaker}, Graph Neural Network (GNN) can effectively leverage relational information between data points to further refine the speaker embeddings, thereby enhancing their representativeness. Wang et al. \cite{wang2020speaker} initially used Graph Convolutional Network (GCN) to refine speaker embeddings, effectively separating different speakers within a conversation. Wang and Zheng \cite{wang2023community,zheng2022reformulating} formulated the clustering task as a community detection problem to achieve speaker classification. Despite significant progress in clustering, these methods still face challenges with overlapping audio, indicating that spatial relationships in speaker embeddings remain underutilized.
\begin{figure}[t]
  \centering
  \includegraphics[width=\linewidth,height=5cm]{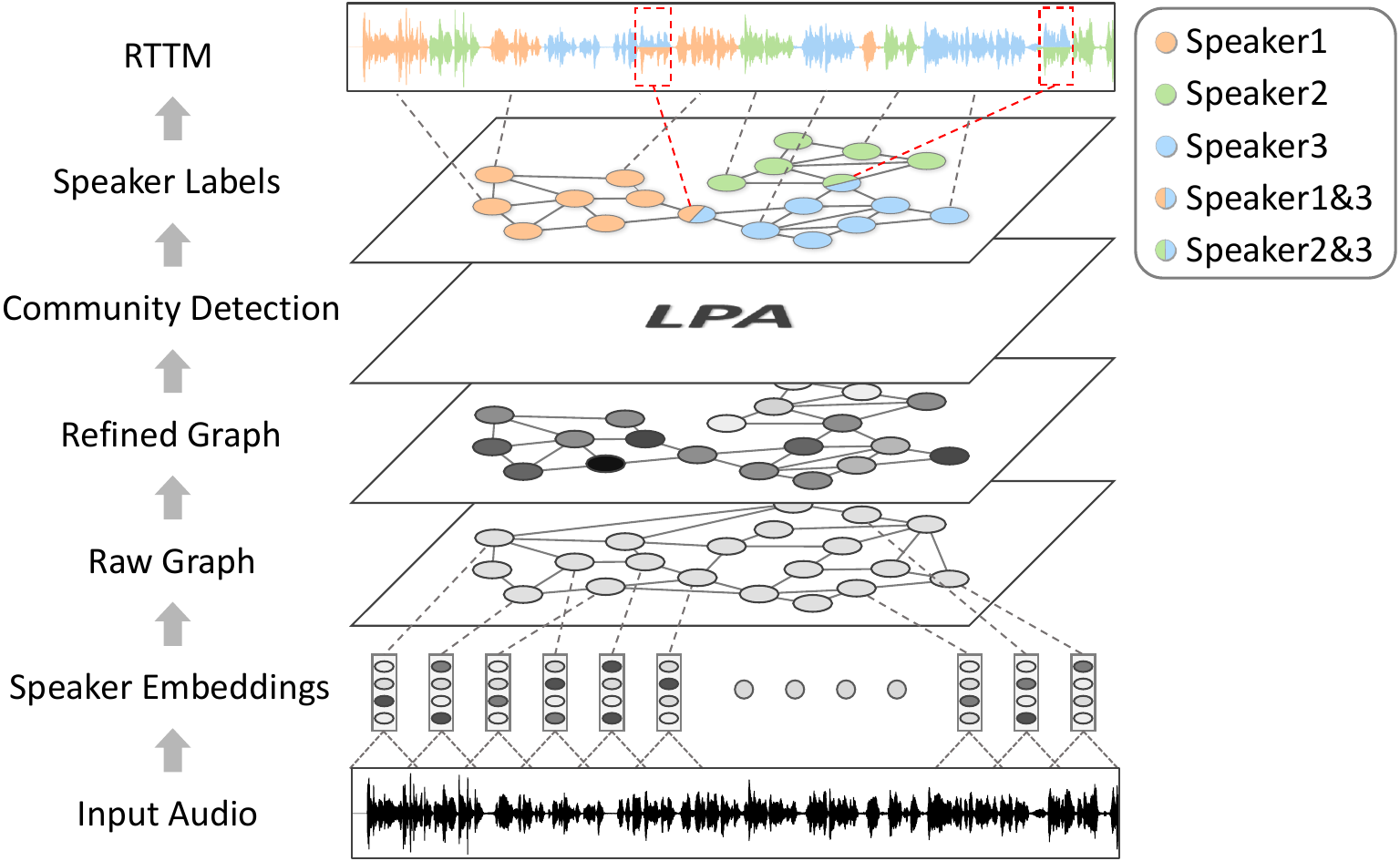}
  \caption{An illustration of the speaker diarization system pipeline with the OCDGALP clustering method.}
  \label{fig:OCDGALP_pipeline}
  \vspace{-2em} % 适当调整数值
\end{figure}

Inspired by the aforementioned works, this paper proposes a novel speaker diarization method based on Graph ATtention network (GAT) and label propagation, with the system flow illustrated in Figure \ref{fig:OCDGALP_pipeline}. Our main contributions are as follows:
\begin{itemize}
\item Enhanced Speaker Embedding Representation: The speaker embedding graph is fed into the GAT, where high-dimensional embeddings are projected into a lower-dimensional space. This transformation provides a clearer representation of speaker relationships, thus facilitating subsequent clustering.
\item Integrated Clustering and Overlap Detection: By combining clustering with overlap detection, label propagation leverages node connections predicted by the graph neural network. This enables iterative partitioning of overlapping communities, effectively addressing overlapping speech.
\item State-of-the-Art Performance: Experimental results show that the proposed method significantly reduces Diarization Error Rate (DER). Specifically, it achieves a remarkable 15.94\% on the DIHARD-III dataset \cite{ryant2020third} without oracle VAD and an impressive 11.07\% with oracle VAD.
\end{itemize}

\section{Related works}

\subsection{Graph attention network}
\label{subsection:Graph attention network}
In our work, a modified GAT model is used to update the speaker embedding graph. Assuming  \(L\)-layer graph attention network, where \( l \in \{1,2,...,L\} \) represents the layer index, we take the embedding matrix \( \mathbf{H} \) \(\in\mathbb{R}^{K \times D}\), which consists of pre-extracted speaker embeddings, along with the adjacency matrix \( \mathbf{A} \)  \(\in\mathbb{R}^{K \times K}\) as the network inputs, where \( K \) is the number of nodes in the graph and \( D \) is the dimension of the embeddings.

Each layer of the graph attention mechanism performs information aggregation for nodes with the following form:
\begin{align}
    z_{i}{}^{(l+1)}=\sigma(\sum_{j\in\mathcal{N}_{i}}\alpha_{ij}^{(l)}\mathbf{W}^{(l)}z_{j}{}^{(l)}).
    \label{equation:eq1}
\end{align}

Here, \(z_{i}{}^{(l+1)}\) represents the output feature of node \( v_i \) in the embedding matrix \( \mathbf{H} \) at layer \( (l+1) \). The set \( N_i \), which includes the neighboring nodes connected to \( v_i \) (including itself), can be obtained from the adjacency matrix \( \mathbf{A} \). The matrix \( \mathbf{W} \) \(\in\mathbb{R}^{D_1 \times D_2}\) represents the trainable weight matrix at layer \( l \), with input and output dimensions of \( D_1 \) and \( D_2 \), respectively. The attention coefficient \( \alpha_{ij} \) determines the importance of the neighboring node \( v_j \) to the node \( v_i \). In addition, \( \sigma \) is a nonlinear activation function. According to the method proposed by Velickovic et al. \cite{velickovic2017graph}, the attention coefficient \( \alpha \) is calculated as follows:
\begin{align}
    \alpha_{ij}&=\mathrm{softmax}_{j}(\widehat{\alpha}_{ij})\nonumber\\
    &=\frac{\exp\left(\delta\left(\mathbf{a}^{T}[\mathbf{W}z_{i}\parallel \mathbf{W}z_{j}]\right)\right)}{\sum_{b\in\mathcal{N}_{i}}\exp\left(\delta\left(\mathbf{a}^{T}[\mathbf{W}z_{i}\parallel \mathbf{W}z_{b}]\right)\right)}.
    \label{equation:eq3}
\end{align}
where \( \mathbf{a} \in \mathbb{R}^{2 \cdot D_2} \) is a learnable weight vector, \( T \) denotes transposition,\(\ ||\) is the concatenation operation, and \( \delta(\cdot) \) is the LeakyReLU activation function.

\subsection{Label propagation algorithm}
\label{subsection:Label propagation algorithm}
The Label Propagation Algorithm (LPA) \cite{raghavan2007near} is a fast and efficient community detection method but struggles with detecting overlaps and maintaining stable results. To overcome these drawbacks, several improved LPA-based methods have been proposed, such as COPRA \cite{gregory2010finding}, SLPA \cite{xie2011slpa}, and DLPA \cite{sun2015detecting}. In our work, we employ a variant of COPRA, known as LPANNI (Label Propagation Algorithm with Neighbor Node Influence) \cite{lu2018lpanni}, whose algorithmic steps are as follows:
\begin{enumerate}
\item For a given network \(G=(V,\mathcal{E})\), where $V$ represents the set of nodes and $\mathcal{E}$ represents the set of edges, and \( u \in V \),  \(v\in Ng(u)\), and \( Ng(u) \) represents the set of neighbors of \( u \), we introduce additional definitions for nodes in the graph: Node Importance (\(\mathit{NI}(u)\)), Node Similarity (\(\mathit{Sim}(u, v)\)), and Neighbor Node Influence (\(\mathit{NNI}_v(u)\)), which are defined in the same way as suggested by Lu et al. \cite{lu2018lpanni}.
\item Nodes are sorted according to \(\mathit{NI}(u)\) for label updating. At this stage, nodes will receive multiple dominant labels from neighboring nodes, forming a label set:
\begin{align}
    L_{Ng}=\{l(c_1,b_1),l(c_2,b_2),...l(c_v,b_v)\},
\end{align}
where $l(c_v, b_v)$ denotes the dominant label of neighbor node $v$, and $b_v$ denotes the belonging coefficient of node $v$ to community $c_v$. The dominant label is the one in the label set with the highest belonging coefficient.
\item Recalculate the new belonging coefficient $b^{\prime}(c,u)$ of node $u$ to community $c$ using $\mathit{NNI}$ and $L_{Ng}$:
\begin{align}
    b^{\prime}(c,u)=\frac{\sum_{l(c_v,b_v)\in L_{Ng},v\in Ng(u),c_v=c}b(c_v,v)NNI_v(u)}{\sum_{l(c_v,b_v)\in L_{Ng},v\in Ng(u)}b(c_v,v)NNI_v(u)}
\end{align}
Then, the updated label set $L^{\prime}$ of node $u$ is generated:
\begin{align}
    L^{\prime}=\{l(c_1,b_1^{\prime}),l(c_2,b_2^{\prime}),...l(c_{|L^{\prime}|},b_{|L^{\prime}|}^{\prime})\},
\end{align}
where $\lvert L^{\prime} \rvert$ denotes  the number of the labels in $L^{\prime}$. Next, labels satisfying \( b^{\prime}(c,u) < 1/\lvert L^{\prime} \rvert \) are adaptively removed, forming the refined label set $\mathit{L''}$.
\item Normalize the belonging coefficients of the labels in $\mathit{L''}$ to obtain the final label set $L_u$.
\end{enumerate}

The update process follows a historical label priority strategy to reduce propagation randomness. If multiple labels share the highest belonging coefficients, the dominant label is chosen from those present in the previous iteration; otherwise, it is selected randomly.

In summary, the LPANNI algorithm enhances the stability of label propagation through the proposed strategy and effectively assigns multiple labels to overlapping community nodes. These characteristics make LPANNI particularly well-suited for addressing overlapping speakers in speaker diarization tasks.

\begin{figure*}[t]
  \centering
  \includegraphics[width=16cm,height=7.5cm]{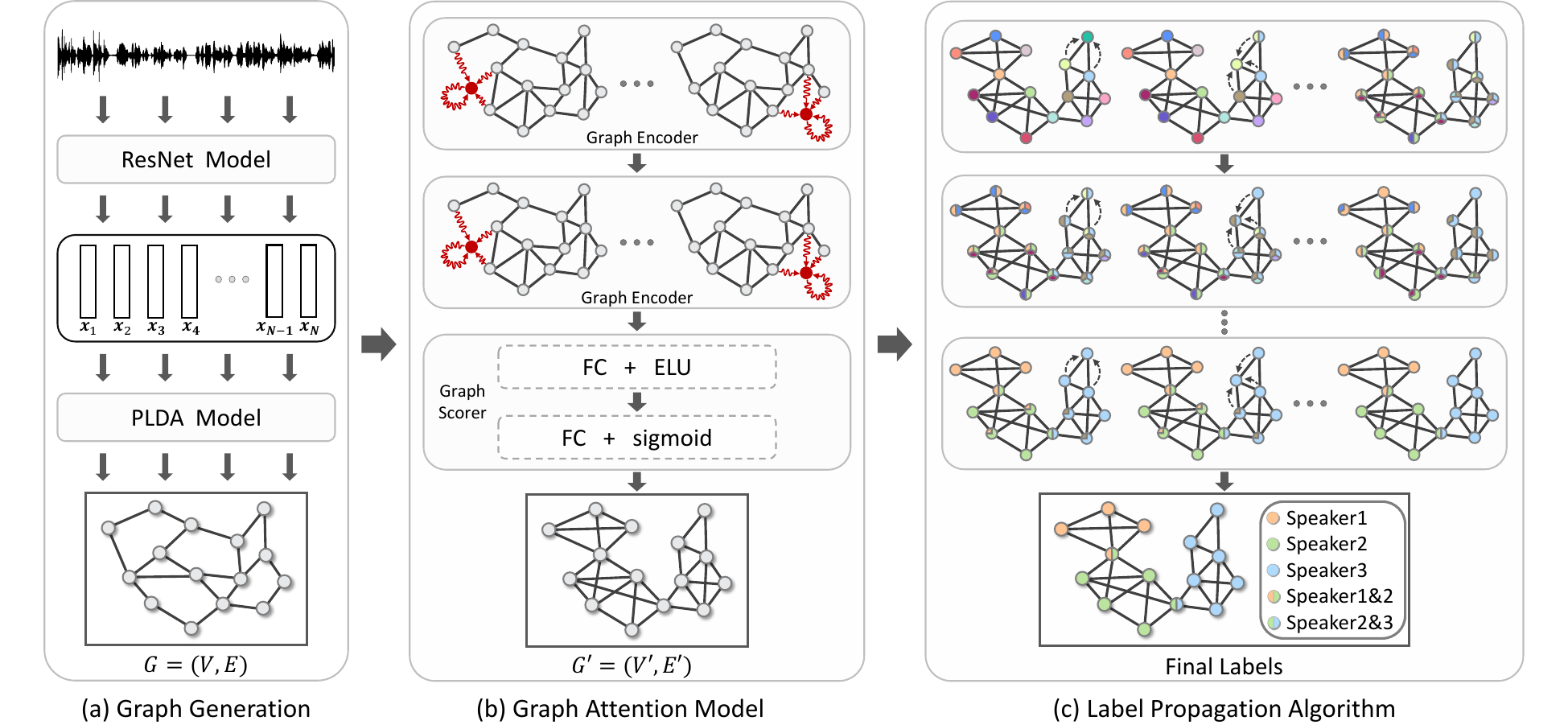}
  \caption{The architecture of the Overlapping Community Detection based on Graph Attention and Label Propagation Algorithm.}
  \label{fig:OCDGALP_architecture}
  \vspace{-1.7em} % 适当调整数值
\end{figure*}
\section{Proposed Method}
The overall framework of OCDGALP is illustrated in Figure \ref{fig:OCDGALP_architecture}. The following sections provide a detailed introduction to each module of OCDGALP.
\subsection{Graph Generation}
In a graph \( G = (V, \mathcal{E}, A) \) composed of conversations, as illustrated in Figure~\ref{fig:OCDGALP_architecture}(a), the nodes \( V = \{ v_1, v_2, \dots, v_N \} \in \mathbb{R}^N \) are derived from speaker embeddings extracted from audio segments, while the edges \( \mathcal{E} \in \mathbb{R}^{N \times N} \) represent the connections between nodes. The affinity matrix \( \mathbf{A} \in \mathbb{R}^{N \times N} \) stores edge weights based on similarity scores computed using probabilistic linear discriminant analysis (PLDA)~\cite{ioffe2006probabilistic}, normalized to the range \([0,1]\). To reduce complexity, only edges with weights above a threshold \( \mu \), treated as a hyperparameter, are retained.

\subsection{Graph attention model design}
As shown in Figure \ref{fig:OCDGALP_architecture}(b), the entire network consists of two components: a graph attention encoder and a scoring module. When the original graph passes through the GAT encoder, it learns information from surrounding nodes with different weights, generating a more refined latent speaker embedding graph. Each pair of nodes in this refined graph is then used as input to the scoring module. 

The scoring module consists of fully connected layers with nonlinear activation functions and is jointly trained with the GAT layers. During this process, it re-predicts the relationships between updated graph nodes to generate the reconstructed speaker embedding graph \( G' \) and the affinity matrix \(\hat{\mathbf{A}}\). To preserve important structural information from the original embeddings during the encoding process, we propose fusing the reconstructed affinity matrix \(\hat{\mathbf{A}}\) with the initial affinity matrix \(\mathbf{A}\). The fused affinity matrix \(\bar{\mathbf{A}}\) is computed as follows:  
\begin{align}
    \bar{\mathbf{A}}=(1-\varepsilon)\hat{\mathbf{A}}+\varepsilon\mathbf{A}
\end{align}
where \( \varepsilon\in[0,1] \) is a balance coefficient that controls the influence of different affinity matrices. The loss function used for model training is defined as the difference between the fused affinity matrix and the ground truth adjacency matrix, calculated using Binary Cross-Entropy (BCE) loss.
\subsection{Overlapping community detection}

As shown in Figure \ref{fig:OCDGALP_architecture}(c), we integrate clustering with overlapping speaker detection and collectively refer to it as overlapping community detection.

The algorithm consists of two main stages. First, it calculates and sorts the \(\mathit{NI}\), primarily to identify key speaker labels as propagation sources while also enhancing the algorithm's stability. Then, we consider the order of \(\mathit{NNI}\) and \( \mathit{Sim} \) to be crucial for node updates. This is particularly important for overlapping nodes, which may be located at community boundaries and are more likely to possess multiple label assignments.

The convergence condition is met when the size of the label set and the dominant labels of all nodes remain stable or when the maximum number of iterations is reached. Upon completion of the iteration \(\tau\), nodes with multiple speaker labels are considered overlapping nodes, and the overlapping community structure is detected. It is worth noting that when computing \(\mathit{Sim}\), a path length threshold \(\beta\) is used to control computational complexity. Both \(\beta\) and the maximum iteration count \(\tau\) are regarded as hyperparameters.

\section{Datasets and experimental setup}
\subsection{Evaluation and training data preparation}
We evaluated our speaker diarization system on the DIHARD-III corpus, which includes a 34.15-hour development (DEV) set and a 33.01-hour evaluation (EVAL) set.
The training sets for different modules in our system are described as follows:
\begin{itemize}
\item \textbf{VAD}: We fine-tuned a suitable VAD model on the DIHARD-III DEV set using the pre-trained models provided by pyannote.audio v3.1\cite{plaquet2023powerset}. In addition, we also conducted experiments using the oracle VAD for comparison.
\item \textbf{Embedding extractor and PLDA training}: We trained the embedding extractor and PLDA model using the VoxCeleb2 dataset, which contains over one million utterances.
\item \textbf{GAT}: We generated a training dataset for the GAT model using VoxCeleb1, VoxCeleb2, and LibriSpeech. Following the methodology from \cite{fujita2019end}, we simulated 6,000 mixed audio recordings featuring 2 to 9 speakers, with durations ranging from 20 to 600 seconds, totaling nearly 1,200 hours of data.
\end{itemize}
\begin{table*}[t]
  \caption{We evaluate the performance of different diarization systems in terms of DER (\%) on the DIHARD III dataset under a 0ms collar condition, which includes the assessment of overlapping speech. The evaluation is conducted on the full dataset, comparing systems both with and without oracle VAD.
}
  \label{tab:diarization_results}
  \centering
  \begin{tabularx}{\textwidth}{
      >{\centering\arraybackslash}m{1.5cm}  % ID 居中
      >{\centering\arraybackslash}m{3cm}  % System 居中
      >{\centering\arraybackslash}m{4.5cm}  % Methods 标题居中
      *{4}{>{\centering\arraybackslash}X}  % Dev/Eval 居中
    }  
    \toprule
    \multirow{2}{*}{\textbf{ID}} &  
    \multirow{2}{*}{\textbf{System}} &  
    \multirow{2}{*}{\makecell{\textbf{Methods}}} &  % Methods 标题居中
    \multicolumn{2}{c}{\textbf{wo/ Oracle VAD}} & 
    \multicolumn{2}{c}{\textbf{w/ Oracle VAD}} \\ 
    \cmidrule(lr){4-5} \cmidrule(lr){6-7}
    &  &  & \textbf{Dev} & \textbf{Eval} & \textbf{Dev} & \textbf{Eval} \\
    \midrule
    S1 & \multirow{3}{*}{\centering Baselines} &
    \makecell[l]{ResNet+PLDA+AHC} & 23.5  & 23.0  & 19.94 & 18.9  \\
    S2 & & \makecell[l]{ResNet+PLDA+SC}  & 22.76 & 22.94 & 17.89 & 16.81 \\
    S3 & & \makecell[l]{Raw-LPA}         & 23.68 & 23.73 & 20.45 & 19.32 \\
    \midrule
    S4 & \multirow{3}{*}{\centering GCN-based} &
    \makecell[l]{ResNet+GCN+AHC}  & 22.36 & 22.45 & 18.04 & 17.55 \\
    S5 & & \makecell[l]{ResNet+GCN+SC}   & 19.89 & 19.94 & 14.03 & 13.12 \\
    S6 & & \makecell[l]{ResNet+GCN+LPA-OCD}  & 16.25 & 16.29 & 11.79 & 11.83 \\
    \midrule
    S7 & \multirow{3}{*}{\centering GAT-based} &
    \makecell[l]{ResNet+GAT+AHC}  & 21.23 & 21.31 & 17.51 & 17.12 \\
    S8 & & \makecell[l]{ResNet+GAT+SC}   & 19.56 & 19.63 & 13.88 & 13.01 \\
    S9 & & \makecell[l]{ResNet+GAT+LPA-OCD}      & \textbf{15.87} & \textbf{15.94} & \textbf{11.03} & \textbf{11.07} \\
    \bottomrule
  \end{tabularx}
  \vspace{-0.3cm}
\end{table*}

\subsection{Experimental setup}
In our diarization systems, we split the audio into 1.5s length segments with 0.75s window shift, and extract 256-dimensional embeddings of segments with the ResNet-34-SE model from ASV-Subtools \cite{tong2021asv}. The model consists of two GAT layers, followed by two fully connected layers with activation functions. The dimensionality settings are [d-128-64-64-1], with ELU and sigmoid as activation functions. The graph construction threshold \( \mu \) is set to 0.3, and the balance coefficient \(\varepsilon \) is set to 0.5.

During the subsequent overlapping community partitioning, we set the maximum number of iterations \( \tau \) to 80 and the path length \( \beta \) to 2. To mitigate the impact of algorithm instability on the results, each experimental outcome is obtained by averaging over 10 repeated trials.

\section{Experimental results}
\subsection{Choice of hyperparameters}
\begin{figure}[t]
  \centering
  \includegraphics[width=\linewidth,height=5.5cm]{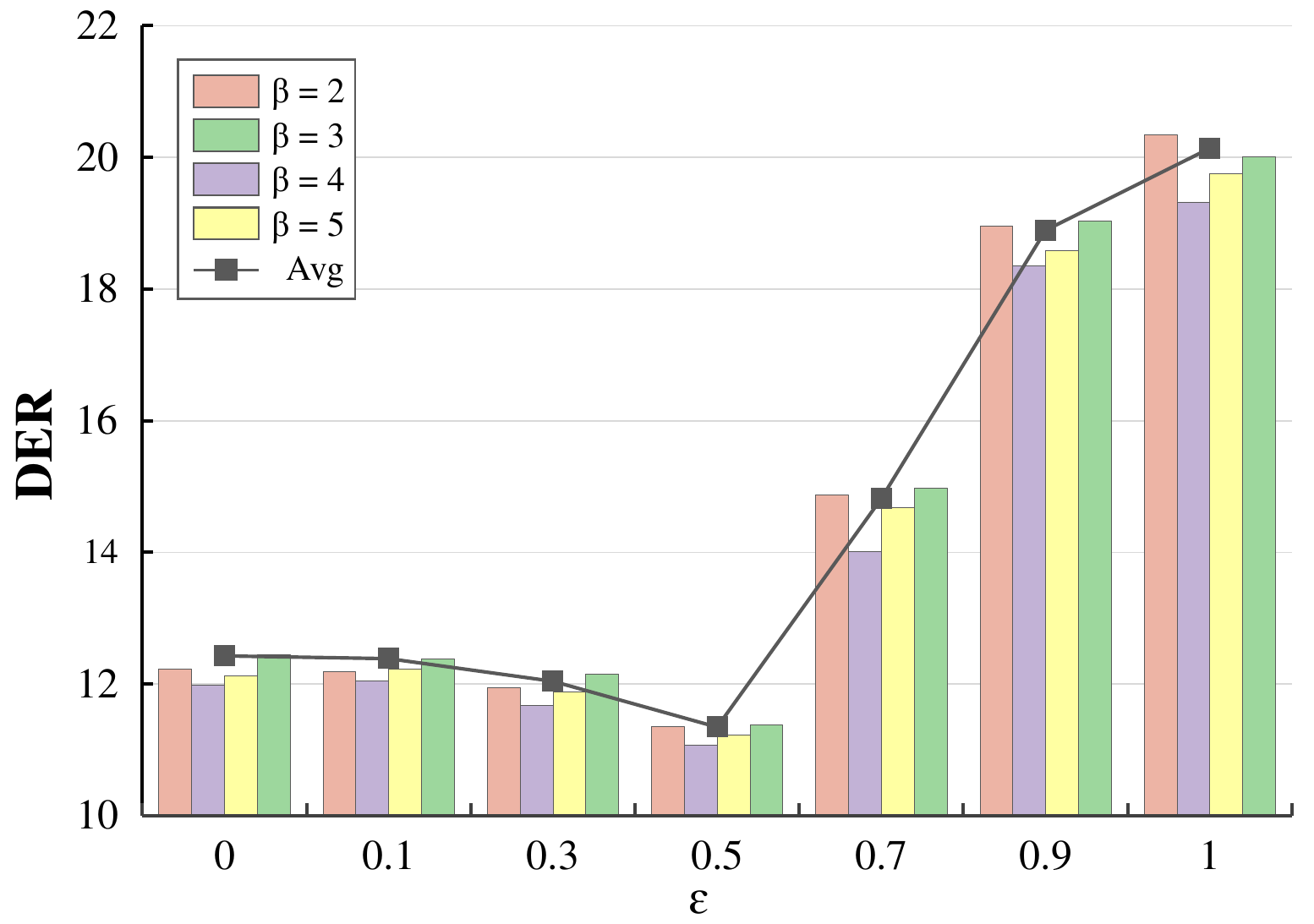}
  \caption{The comparison of DER(\%) with different \(\varepsilon \) and \( \beta \) on the DIHARD III dataset.}
  \label{fig:hyperparameters}
  \vspace{-2em} % 适当调整数值
\end{figure}

We investigate the effect of the balance coefficient \( \varepsilon \) and the path length threshold \( \beta \) on the OCDGALP system. As shown in Figure \ref{fig:hyperparameters}, the lowest DER is obtained when \( \varepsilon = 0.5 \) and \( \beta = 3 \). This result demonstrates the effectiveness of using a fused affinity matrix and suggests that an optimal path length is crucial for efficient label propagation, as excessively long or short paths may hinder performance.
\subsection{Speaker clustering results}

In this section, we examine the impact of different modules on the performance of the speaker diarization system. We define three types of systems for comparison: S1 to S3 serve as baseline systems, each employing a different clustering method—AHC, SC, and LPA, respectively. In contrast, S4 to S6 integrate GCN modules, while S7 to S9 incorporate GAT modules into the baseline systems. Notably, S9 represents our proposed OCDGALP system.

As illustrated in Table \ref{tab:diarization_results}, the GAT-based system consistently outperforms both the baseline and GCN-based systems across all clustering methods. This highlights the superior effectiveness of the proposed GAT module. The attention mechanism in the GAT module, which assigns weighted importance to relationships between different nodes, is more effective than the uniform aggregation of node information in GCN. Notably, although Raw-LPA performs worse than AHC and SC when used independently, it significantly outperforms  other clustering methods when integrated into the updated graph structure with GCN and GAT. This improvement is primarily attributed to its ability to detect overlapping structures, which enhances the overall performance of the proposed system.
\subsection{Performance comparison}

As shown in Table \ref{tab:COMPARISON_results}, we compare our proposed OCDGALP-based system with the latest systems in the literature on the DIHARD III dataset. To ensure a fair evaluation, we adhere to the same conditions specified in the original challenge: no collar tolerance is applied, and overlapping speech is considered.

The results clearly demonstrate that our system, when equipped with oracle VAD, not only outperforms existing single-clustering systems but also matched the performance of the best system reported to date. Furthermore, in scenarios without oracle VAD, our system achieves the best publicly available performance with a DER of 15.94\%, further demonstrating the effectiveness of our OCDGALP model.

\begin{table}[t!]
    % \flushright
    \centering
    \caption{DER(\%) on the DIHARD III: A comparison with other systems in the literature}
    \label{tab:COMPARISON_results}
    
    \begin{tabular}{l l c}
        \toprule
        \textbf{System} & \textbf{Method} & \textbf{DER(\%)} \\
        \midrule
        \multicolumn{3}{l}{\textbf{With Oracle VAD}} \\
        \hspace{5pt} SA-EEND\cite{fujita2019end} & EEND & 16.19 \\
        \hspace{5pt} EEND-EDA\cite{horiguchi2022encoder}  & EEND & 14.42 \\
        \hspace{5pt} Hitachi /- fusion\cite{horiguchi2021hitachi} & Fusion/cluster & 11.58 / 15.74 \\
        \hspace{5pt} USTC System\cite{wang2021ustc} & Fusion & 11.3 \\
        \hspace{5pt} BUT system/- fusion\cite{landini2021but} & Fusion/cluster & 14.56 / 15.65 \\
        \hspace{5pt} ANSD-MA-MSE\cite{he2023ansd} & TS-VAD & 11.12 \\
        \hspace{5pt} Seq2Seq-TSVAD\cite{cheng2023target} & TS-VAD & \textbf{10.77} \\
        \hspace{5pt} OCDGALP(ours) & Cluster & \underline{11.07} \\        
        \midrule
        \multicolumn{3}{l}{\textbf{Without Oracle VAD}} \\
        \hspace{5pt} Pyannote.audio v3.1\cite{plaquet2023powerset}  & Cluster & 21.3 \\
        \hspace{5pt} EEND-EDA\cite{horiguchi2022encoder}  & EEND & 20.69 \\
        \hspace{5pt} DiaPer\cite{landini2024diaper} & EEND & 20.3 \\
        \hspace{5pt} EEND-GLA\cite{horiguchi2022online} & EEND & 19.49 \\
        \hspace{5pt} VBx + Reseg\cite{bredin2021end} & Cluster & 19.3 \\
        \hspace{5pt} USTC System\cite{wang2021ustc} & Fusion & 16.78 \\
        \hspace{5pt} ANSD-MA-MSE\cite{he2023ansd} & TS-VAD & 16.76 \\
        \hspace{5pt} EEND-M2F\cite{harkonen2024eend} & EEND & \underline{16.07} \\
        \hspace{5pt} OCDGALP(ours) & Cluster & \textbf{15.94} \\
        \bottomrule
    \end{tabular}
    \vspace{-0.6cm}
\end{table}
\section{Conclusion}
In this paper, we propose the OCDGALP model for speaker diarization. Our primary goal is to leverage the topological structure of speaker embeddings and fully exploit graph-based information to achieve both clustering and overlap detection. To this end, we incorporate a GAT model to optimize the graph structure formed by speaker embeddings. Additionally, we employ LPA to determine the optimal partition of overlapping communities. Experimental results demonstrate that the OCDGALP-based speaker diarization system outperforms the baseline and several recent speaker diarization systems on the DIHARD-III corpus when oracle VAD is used. Moreover, it achieves state-of-the-art performance even without oracle VAD.
\section{Acknowledgement}
This work was supported in part by the National Natural Science Foundation of China under Grants 62371407 and 62276220, and the Innovation of Policing Science and Technology, Fujian province (Grant number: 2024Y0068)
\bibliographystyle{IEEEtran}
\bibliography{main}

\end{document}